\documentclass{article}%
\usepackage{amsmath}
\usepackage{amsfonts}
\usepackage{amssymb}
\usepackage{graphicx}
\usepackage{multirow}

\newcommand{\MH}{Michel H\'enon}

\title{\MH, a playfull and simplifying mind}
\author{Jean-Marc Petit\\
\small Institut UTINAM - CNRS - Universit\'e de Franche Comt\'e\\[-0.8ex]
\small 41 bis avenue de l'Observatoire, F-25010 Besan\c{c}on cedex, FRANCE\\
\small \texttt{Jean-Marc.Petit@normalesup.org}
}
\date{\today}

\begin{document}
\maketitle


\section{Introduction}

Several chapters in this book present various aspects of {\MH}'s
scientific acheivements that spread over a large range of subjects, and yet
managed to make deep contributions to most of them. The authors of these
chapters make a much better job at demonstrating the big advancements that
{\MH} allowed in these fields than I could ever do. Here I rather present some
facets of his personnality that most appealed to me.
{\MH} was a reserved person, almost shy, so it was not obvious for a young
student to grasp the profoundness of his insight and what a marvelous advisor
he could be.
The two most prominent aspects of his mind, in my view, were his ability to
simplify any scientific question to its core complexity, and to find the fun
and amusing part in his everyday work, even in the tiniest details of his
scientific investigations.

\section{My first meeting with Michel H\'enon}

{\MH} was a teacher in the Dipl\^ome d'\'Etude Approfondie (DEA, the equivalent
of a Master at the time in France) {\it Turbulence et Sys\`emes
  Dynamiques}. Our first class with him was not on the study of dynamical
systems, despite all his contributions to the field. Rather, it was a {\it
  practical} lecture on programming. Other teachers in this master were in
charge of teaching us the theory and analytical study of dynamical
systems. Since he professed to no be a good mathemacian nor theorist, which
none of us believed, he was concerned with numerical studies of dynamical
system. Thus he felt that we needed to be tought basic good practice in
programming to make sure we could the problem with the correct tools.

{\MH} was pragmatic and wanted to teach us useful knowledge. He could have
decided to teach us an object-oriented language for software scientist that is
very strict ...
Instead, he decided to use the main language used in computational physics,
Fortran, in its current incarnation Fortran77, and to give us very simple
recipes to make our programmes clear, readable, correct and resilient to basic
errors such as typos. He relied on the good will of the programmers rather than
on the grammar of the language and the compiler to write \textit{good}
programmes.

In {\MH}'s view, one needs to have a structured mind to address any physical or
mathematical question, and then one must have a structured view of the problem
at hand. It followed that the tools used to solve the problem, in this case a
computer programme, had to be well structured.

\begin{figure}
\hspace{-1.5truecm}
\begin{minipage}{1.0\textwidth}
\begin{minipage}{0.6\linewidth}
(a) \\
\includegraphics[width=6.5cm, angle=-90]{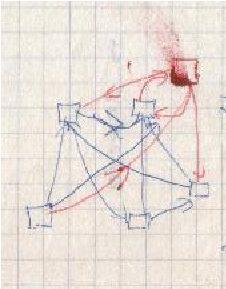} 
\end{minipage}
\hspace{0.2truecm}
\begin{minipage}{0.6\linewidth}
(b) \\
\includegraphics[width=6.5cm, angle=-90]{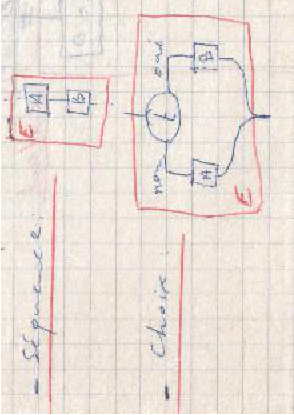} 
\end{minipage}
\\ [0.4cm]
\begin{minipage}{0.6\linewidth}
(c) \\
\includegraphics[width=6.0cm, angle=-90]{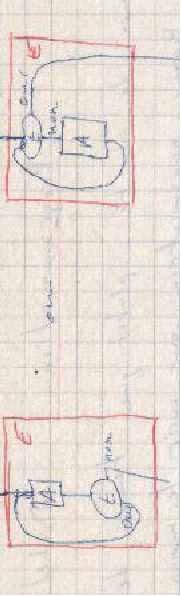} 
\end{minipage}
\end{minipage}
\\ [0.4cm]
\caption{(a): the too frequent spaghetti plate type of programming, or
mess, that should be avoided. Instead, programmes should be built from simple
blocks. Only a small number of block types are needed to construct any
programme, like sequences, tests (b), or loops (c). These graphs are copies of
my personnal notes at the time, which reproduced as best as I could the drawing
from {\MH} on the black board.}
\label{fig:progs}
\end{figure}

The main concern when writing a computer programme should be clarity, over
computing time and memory ressources. A good organisation is instrumental in
avoiding errors from the start, thus limiting the possibility of uncaught
errors at the end. If $L$ is the length of the programme (say number of lines
or statements), then the tuning time for a spaghetti-like programme (see
Fig.~\ref{fig:progs}) goes like $L^2$, while that of a well structured
programme is proportional to $L$. Since human time is more valuable than
machine time, a programme has to be well structured. To achieve this, a
programme should be organised using modules and sub-modules. As a bonus, it
turns out that in most cases, this will allow the compiler to produce a more
efficient binary code.

Structured programmation is based on blocks. A block has a single input point
and a single output point. Blocks can be compsed together to form bigger
blocks. One can define three different types of basics blocks: a sequence, an
test or choice, and a loop. Obviouly there are variants of these blocks. A test
can have only two possible outcomes (say
\verb|if (x < 0) then {...} else {...}|) or a whole complement of
possibilities (like in the \verb|case ...| statement). A loop can have the
completion test at the star (\verb|do while (test) {...}|), at the end
(\verb|repeat {...} until (test)|) or in the middle
(\verb|mark {...} (test) {...} goto mark|). Equipped with these tools, {\MH}
showed us how to use them on a simple example, counting the number of cycles in
a bijection mapping on a group of 6 elements. For us, fresh master students who
had barely had any programming class, this example was enlightening.

The structure of the programme can and should also be reflected in its actual
appearance on the screen or listing. Within the limitations of the language
grammar, one should choose variable and subroutine names that are meaningful.
The text should be indented to reflect the structure of the logical blocks and
modules. (Interestingly, this approach was pushed to the extreme in a very
sucessful language, Python, where the scope of a block is given by indentation
of the statement lines.) Because our mind get a better grasp of what we see at
once, one should limit the size of a module to the size of a page or screen, by
using sub-modules when possible.

As important as the structure of the programme, one must document it. Comments
should appear everywhere in the programme. They are of utmost important for
long term maintenance of the code. At the start of the code, one should write
the details of the problem at hand, give the equations solved by the programme,
give a list of all the variables (and always declare the type of the variable,
whether this is mandatory for the programming language or not), and of the
subroutine (modules) and their purpose. Together with this detailed
description, one should also include the user's manual. Typical of his way of
doing things, {\MH} wrote a frotran programme that would take another fortran
code with comments written in TeX format and produce a well written paper with
comments forming the core text, and the computer code written verbatim in
between.

Straight from the beginning, and through this simple class, {\MH} managed to
convey a set of very important rules:
\begin{itemize}
\item Start with a global vision, apply a \textit{top-down} approach;
\item Extract the hard point of the problem;
\item Perform a structured analysis of the problem;
\item Write a clear documentation
\item Apply a rigorous methodology.
\end{itemize}
The main strength of {\MH} was to stick to these rules by all means.

While teaching his class, {\MH} was calm, quiet and reserved. He really
focussed on the essence of what he wanted to tell us. He avoided unnecessary
complications aimed at showing how clever he was. This simplified and rigorous
approach made a strong impression on the students and appealed very much to me.
His very quiet style made him stand out amongst the teachers of the DEA.

\section{Saturn's rings}

I had to do a DEA reasearch project in spring 1983, hoping to continue on a 3rd
cycle PhD thesis (the shorter format of a PhD thesis that was current at the
time in France), at the time when we got the first Voyager data on Saturn's
rings.

\subsection{Context of our work}
The rings of Saturn have an almost perfect circular symmetry; they are also
extremely flat. Deviations of particle orbits from circular and coplanar
shapes are of the order of $10^{-6}$, with a few exceptions (eccentric rings,
irregular rings, spokes). Therefore their spatial structure is essentially
described by a single function: the {\sl radial distribution}. Until that time,
this distribution was believed to be also rather simple and smooth. The few
structural details which could be seen from the Earth were considered as
remnants of the formation process, or, as in the case of the Cassini division,
were attributed to resonances with the major satellites.

\begin{figure}
\hspace{-1.5truecm}
\begin{minipage}{0.6\linewidth}
(a) \\
\includegraphics[width=5.5cm, angle=-90]{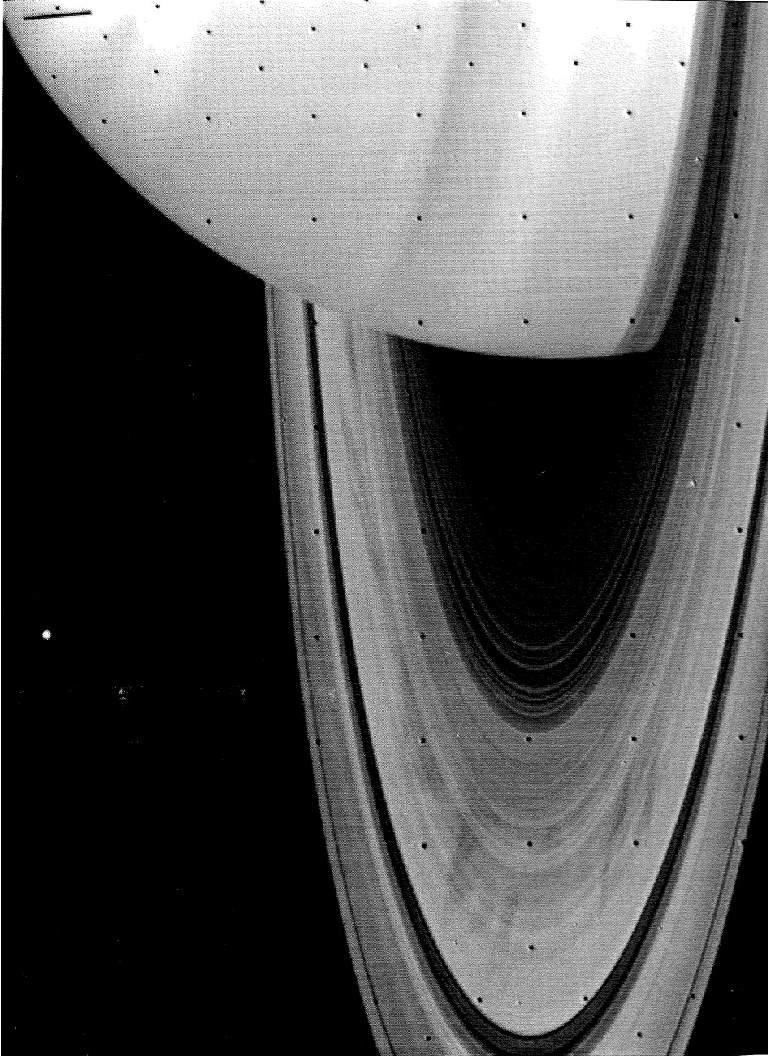} 
\end{minipage}
\hspace{0.0truecm}
\begin{minipage}{0.6\linewidth}
(b) \\
\includegraphics[width=7.0cm, angle=90]{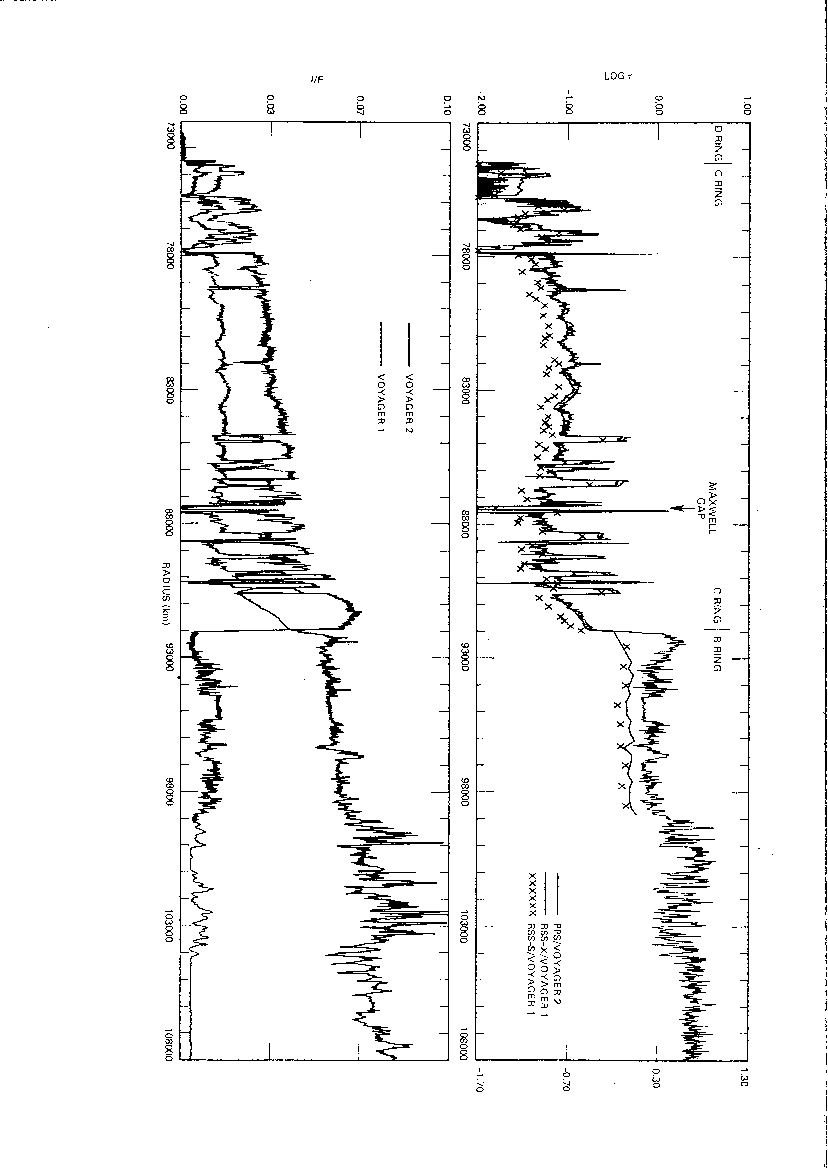} 
\end{minipage}
\\ [0.4cm]
\caption{(a): One of the first photos of Saturn's rings taken by the
  interplanetary probe VOyager 1. (b): 
}
\label{fig:rings}
\end{figure}

The observations made by the Voyager probes have shattered this last belief and
have revealed that the radial distribution is in fact extremely complex, with
structure at all wavelengths down to the limit of resolution
(Fig.~\ref{fig:rings}). On the other hand, the circularity and flatness of the
rings have been confirmed and even sharpened.

A major challenge for theorists was of course to explain these radial
distributions. Essentially three kinds of explanations had been advanced: (i)
resonances with external satellites; (ii) collective effects, leading to
instabilities; (iii) cumulative effect of binary interactions. However, there
were difficulties with each of these approaches, and at the time it is not
clear which theory, or combination of theories, would ultimately provide the
correct explanation.

{\MH} addressed the problem in his typical simplifying way (Petit \& H\'enon,
1987):

{\it ``In the present state of our knowledge, it seems premature to try to 
set up a fully realistic model of the rings...
Therefore our objective will be, more modestly, to try to gain an 
understanding of some of the fundamental mechanisms at work.
We will include in the model only some selected effects, and ignore 
the others. 
Thus, our ring models should be thought of as
``model problems''. Our hope is that they will behave in 
some fundamental respects like real rings, and thus teach us something 
about ring physics.''}

We thus constructed a \textit{simplified} that retained the essence of the
problem.
\begin{itemize}
\item We consider a 2-dimensional problem;
\item The evolution of the system is a succession of binary interactions;
\item Two physical effects of equal importance are included in the interactions:
  gravitation and inelastic collisions;
\item Since we are interested in the radial structure, keep only the radial
  coordinate in the global evolution;
\end{itemize}

\subsection{Satellite encounters and computer algebra}

Once this course of action was defined, we first had to determine the effect of
gravitational and collisional interaction between two particles in orbit around
Saturn, at large distance. On the gravitational part, this is exactly the
Hill's problem, first defined to study the motion of the Earth-Moon system
around the Sun. Hill's equations are usually derived assuming a hierarchy of
masses for the three bodies:
\begin{equation}
m_1 \gg m_2 \gg m_3,
\end{equation}
an then proceeding in two steps: first take the limit $m_3 \to 0$, which gives
the restricted three-body problem; then take the limit $m_2 \to 0$. Hill's
problem is thus presented as a sub-case of the restricted three-body problem.

We showed that we can consider a more general situation: the ratio of the two
masses $m_2$ and $m_3$ can be arbitrary; the only condition is the both masses
should be small compared to $m_1$:
\begin{equation}
m_1 \gg m_2, \qquad m_1 \gg m_3.
\end{equation}
So we fix the ratio $m_2/m_3$ and let both $m_2$ and $m_3$ tend to zero
simultaneously. The equations obtained in this limit are identical to the
classical Hill's equations, showing that (H\'enon \& Petit, 1986):

{\it ``The restricted problem is applicable to situations where one mass is much
smaller than the two others; Hill's problem is applicable to situations where
one mass is much larger than the two others.''}

Once we have the Hill's equations of motion for the relative motion of the two
satellites (masses $m_2$ and $m_3$), we must resort to numerical integration to
find the motion. But this allows to know the motion only on a finite interval,
where the gravitation between the two satellites plays a role. To determine a
full solution from $t = -\infty$ to $t = +\infty$, we must develop analytic
approximations, in the form of asymptotic series for the solution in the limit
$t \to -\infty$ and $t \to +\infty$, i.e. when the satellites are far from each
other. Let us call $h$ the difference in initial semimajor-axis of the
satellites before interaction, expressed in Hill's coordinates, and $\eta$ the
Hill's coordinate in the direction of relative motion at large distance. In the
asymptotic expansion, the small parameter is $\eta^{-1}$. For initially
circular orbits, we obtain series in powers of $\eta^{-1}$, with coefficients
of $\eta^{-i}$ for $i > 0$ involving powers of $h$ ranging from $h^{i-2}$ to
$h^{-2i+1}$. This is cumbersome to derive, but still doable. 

\begin{figure}
\includegraphics[width=15cm, angle=0]{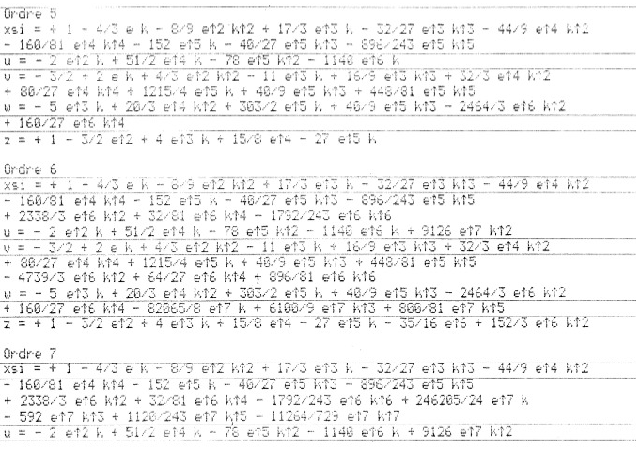}
\caption{Printer output from {\MH}'s computer algebra program applied to the
  Hill' equation asymptotic expansions. Order 5, 6 and part of order 7 in
  $\eta^{-1}$ are displayed.}
\label{fig:algebra}
\end{figure}

But when considering initially eccentric orbits, we have to deal with
trigonometric series in the coefficients of the powers of $\eta^{-1}$,
involving an angle $\theta$ and the relative eccentricity in Hill's
coordinates, $k$. {\MH} could not resist the urge to use a computer to derive
these formulae. One must remember than in the mid-1980's very little was
available in terms of computer algebra, or even easily programmable
computers. {\MH} had acquired a Do-It-Yourself computer like a Zenith, and
decided to use it to derive the series. The computer came with a very simple
OS, a line editor, and a programming language, FORTH. He first wrote a small
program to create a full-screen editor, because he felt writing a large program
with a line editor was not convenient. Once this was available, he wrote a
computer algebra program that could do power series expansions and
trigonometric series expansions. With this tool at hand, he addressed our
asymptotic expansion. One must realise that the full set of tools, and all the
expansion coefficients had to fit inside the 32 KB or memory available in
{\MH}'s computer. The output of this program is shown in
Fig.~\ref{fig:algebra}. Although we were confident the program worked, we still
decided to check it. So each of us independently verified the expansion up to
order 4. The program was right from the beginning, and we finally agreed with
it. The expansions are of the form:
\begin{align}
\xi = & h + k \cos{\theta} - \frac{4}{3} s h^{-1} \eta_c^{-1} +
\left(-\frac{8}{9}h^{-3} + \frac{7}{6}s h^{-1} k \sin{\theta}\right)
\eta_c^{-2} \nonumber \\
 & + \left(\frac{17}{3}s h - \frac{32}{27}s h^{-5} -\frac{7}{3}s h^{-1} k^2 +
\frac{23}{4}s k \cos{\theta} + \frac{28}{27}h^{-3} k \sin{\theta}\right)
\eta_c^{-3} \nonumber \\
 & + 
\left[
-\frac{44}{9} h^{-1} - \frac{160}{81}h^{-7} -\frac{14}{9}h^{-3} k^2 -
\frac{49}{72}h^{-2} k \cos{\theta} \right. \nonumber \\
 & \left. + \left(-\frac{473}{16}h k + \frac{14}{9}h^{-5} k + \frac{99}{32}h^{-1}
k^3\right) \sin{\theta} - \frac{5}{4}s k^2 \sin{2\theta}
\right]
\eta_c^{-4} + O(\eta_n^{-5}),
\label{eq:asymptotic}
\end{align}
where $s = {\rm sign}(\eta)$. The other position and velocity coordinates have
similar expansions.

%

\section{The inclined billiard}

With the previous ingredients, we studied extensively the one-parameter family
of orbits obtained by varying $h$ for initially circular orbits (family of
Satellite Encounters or SE) (Petit \& H\'enon, 1986). This family was found to
be of amazing complexity; in fact it seems to possess the inexhaustible
richness of details which is characteristic of nonintegrable problems in
general.

\begin{figure}[h]
\includegraphics[width=15cm, angle=0]{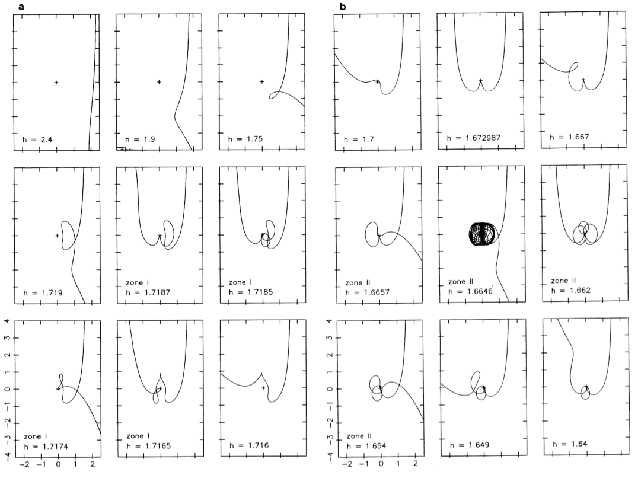}
\caption{(a) Beginning of family SE. Each frame corresponds to one particular
  value of the family parameter $h$. The curve represents the relative motion
  of one satellite with respect to the other, in Hill's coordinates ($\xi$ in
  abscissa, $\eta$ in ordinate). The inital approach is downward from $\eta =
  +\infty$, in the first quadrant. The orbit belonging to transition zones are
labeled zone I, zone II, ... (b) Continuation of family SE.}
\label{fig:SE}
\end{figure}

Figs.~\ref{fig:SE}a and b, taken from a collection of several hundred pictures,
represent the relative motion $(\xi, \eta)$ of the two satellites in Hill's
coordinates. For their description, it will be convenient to think of the
special case $m_2 \gg m_3$, and to identify the origin of the $(\xi, \eta)$
with the satellite $M_2$; the curves represent then simply the motion of
satellite $M_3$.

Three successive phases can be distinguished in a typical orbit: (i) {\it
  approach} of the two satellites; (ii) {\it interplay}, or {\it temporary
  capture:} the two satellites remain close to each other (their distance is of
order 1 in Hill's coordinates) and they perform complex relative motions; (iii)
{\it departure:} the two satellites move away from each other. It can be shown
that {\it permanent capture is possible only for a set of initial conditions of
  measure zero.}

The departure is asymptotically described by (\ref{eq:asymptotic}), with $h$
replaced by $h'$ for the final value, and two cases can be distinguished: (i)
if $h > 0$, then $\eta \to -\infty$, while $\xi$ remains finite and oscillates
around a positive mean value; (ii)if $h < 0$, then $\eta \to +\infty$, and
$\xi$ oscillates around a negative mean value. When $h$ varies, the orbit
alternates from one kind of departure to the other.

For large values of $h$, the orbit of $M_3$ is only slightly perturbed. As $h$
diminishes, the perturbation increases (Fig.~\ref{fig:SE}, $h = 1.9$ and a loop
appears ($h = 1.75$). The shape of the orbit begins to change rapidly with
$h$. Betweene $h = 1.7188$ and $h = 1.7164$ approximately, the orbit undergoes
a series of complex changes of shape. This is the first transition zone (zone I
in Fig.~\ref{fig:SE}). From $h = 1.7164$ to $h = 1.6664$ approximately, things
quiet down and the evolution of the family can again be followed: the shape of
the orbit changes continuously and comparatively slowly with $h$. Then a new
interval of violent changes begins, between $h = 1.6664$ and $h = 1.6497$. This
is the second transition zone (zone II in Fig.~\ref{fig:SE}). More transition
zones occur when $h$ diminishes to zero, separated by quiet intervals.

It should be noted that there is nothing absolute about the limits of the
transition zones, as described above, nor even about their number. When
descending to a finer level of details, one find that each of the transition
zones is resolved into several thinner transition regions, separated by quiet
regions.

\begin{figure}[h]
\begin{minipage}{0.47\linewidth}
\includegraphics[width=7cm, angle=0]{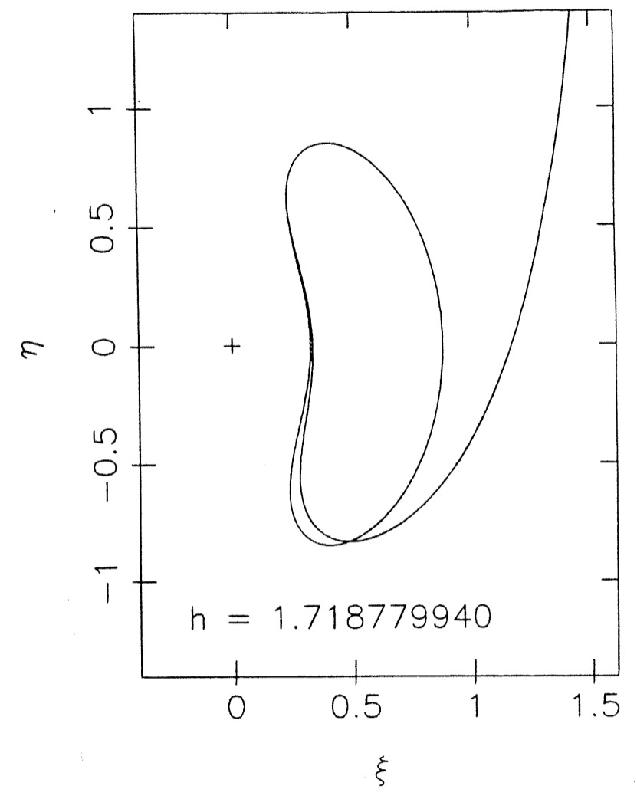}
\caption{An orbit of family SE which is asymptotic to an unstable periodic
  orbit.}
\label{fig:cycle}
\end{minipage}
\hspace{0.4truecm}
\begin{minipage}{0.47\linewidth}
\includegraphics[width=7cm, angle=0]{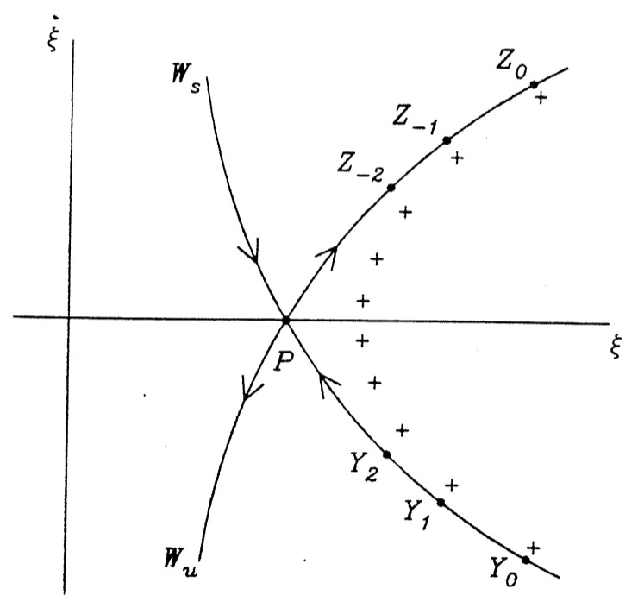}
\caption{Sketch of the surface of section. The value of $\lambda_1$ has been
  artificially increased to show the structure more clearly.}
\label{fig:pointfixe}
\end{minipage}
\\ [0.4cm]
\end{figure}

The net effect of the encounter is essentially characterized by the change in
the final impact parameter $h'$. The transition zones correpond to intervals
where $h'$ changes abruptly from positive to negative value and back with small
changes of $h$. Since $|h'| \ge |h|$, transitions imply a discontinuity in the
family of orbits. This is puzzling since the differential equations that govern
the motion contain no true singularities. Therefore the position of $M_3$ after
a given time should be a continuous function of its initla position and
velocity. To achieve a transition, we must pass through an orbit for which the
duration of the ``temporary capture'' is infinite. This is achieve when {\it
  the orbit tends asymptotically toward a periodic orbit.} This is confirmed
numerically. Fig.~\ref{fig:cycle} represents the orbit for $h = 1.718779940$
which is the first transition encountered when coming from high values of $h$.

To understand what happens, we introduce the surface of section defined by
$\eta = 0$, $\dot\xi > 0$: for each crossing of an orbit with the $\xi$ axis in
the positive direction, we plot a point with coordinate $\xi, \dot\xi$
(Fig.~\ref{fig:pointfixe}). An orbit is represented by a sequence of
points. For a given value of the Jacobi constant $\Gamma$, a point in the
surface of section defines completely the corresponding orbit. In particular,
the next intersection point can be found. This defines the {\it Poincar\'e map}
of the surface of section onto itself.

In the particular Poincar\'e map corresponding to the value of $\Gamma$ for the
orbit in FIg.~\ref{fig:cycle}, the periodic orbit corresponds to the fixed
point $P$ (Fig.~\ref{fig:pointfixe}). This orbit is unstable since it admits an
asymptotic orbit. It has two real eigenvalues $\lambda_1 \simeq 1/640$ and
$\lambda_2 \simeq 640$. An incoming orbit, associated with $\lambda_1$, is
represented by an infinite sequence of points $Y_0$, $Y_1$, $Y_2$, $\dots$,
which lie on the {\it stable invariant manifold of P} converging exponentially
on $P$ (Fig.~\ref{fig:pointfixe}).

An outgoing orbit, associated with $\lambda_2$, corresponds to a sequence of
points $\dots$, $Z_{-2}$, $Z_{-1}$, $Z_0$ lying on the {\it unstable invariant
  manifold} and diverging exponentially from $P$.

The fixed large values of the eigenvalues of the asymptotic periodic orbits
responsible for the transitions preclude a detailed study of this phenomenon.
To study the transition phenomenon in more detail, {\MH} developed a {\it model
  problem} which exhibits basically the same phenomenon and which is easier to
study. With his usual very sharp insight, he decided to consider a problem that
can be reduced to an explicit mapping with smaller and adjustable eigenvalues.

He thus defined the {\i inclined billiard} as follows (H\'enon, 1988). A point
particle moves in the $(X, Y)$ plane. It bounces elastically on two fixed disks
with radius r and with their centres in $(-1, -r)$ and $(1, -r)$. In addition,
it is being subjected to a constant acceleration g which pulis it in the
negative $Y$ direction. Obviously, for most initial conditions, the particle
will after a finite number of rebounds "fall" downwards, never to return. This
is the equivalent of the separation of the two bodies in Hill's problem. To
simplify the numerical computation of the mapping, {\MH} considered the {\it
  large r limit} which replaces the determination of the intersection of a
parabola and a circle with that of two parabolas with parallel asymptotic
directions. In this way, the full orbit can be computed analytically.

He showed that the various parameters defining the problem (total energy $E$,
radius of the disks), can be reduced to a single dimensionless parameter $\Phi$
related to $E$ and $r$.

\begin{figure}[th]
\includegraphics[width=14cm, angle=0]{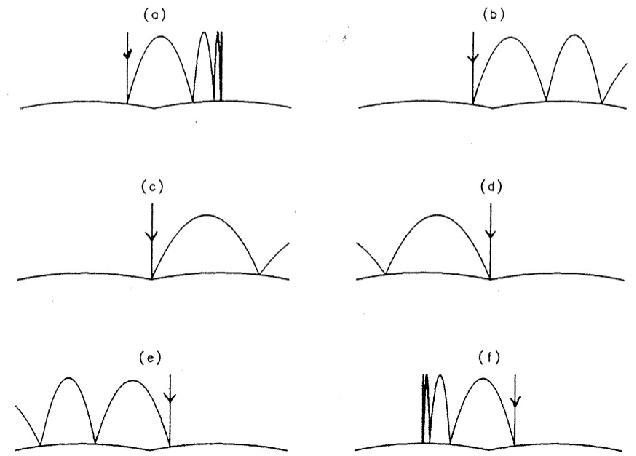}
\caption{Selection of simple members of the $h$-orbit family: (a)
  right-asymptotic orbit; (b), (c) right-escaping orbits ; (d), (e)
  left-escaping orbits; (f) left-asymptotic orbit.}
\label{fig:billiard}
\end{figure}

He then defined a one-parameter family of orbits as in Hill's case, by assuming
that the particle is initially at rest at a position $(h, Y_o)$, where $Y_o$ is
a positive constant and $h$ is variable. This family of $h$-orbits exhibit the
same kind of structure as Hill's problem. We find intervals of continuity, in
which the orbit changes continuously, and transitions. For $h = -1$ , for
instance, the particle bounces indefinitely on the left disk in a straight
vertical line; this is a periodic orbit, which is obviously unstable.  $h = -1$
is a transition value, which separates two quite different kinds of motion: for
$h < -1$, the particle falls toward the left and never returns, while for $h >
-1$ it moves to the right and complex interplays involving the two disks are
possible. A similar periodic orbit exists for $h = +1$. More generally, we may
expect a transition for any solution which is asymptotic to one of these
periodic orbits (Fig.~\ref{fig:billiard}).

Thanks to its careful design, {\MH} was able to derive a symbolic
representation of the $h$-orbits. To a given $h$-orbit, he associated a
sequence of binary digits
\begin{equation}
D: d_1, d_2, \dots, 
\end{equation}
with
\begin{align}
d_j = 
\begin{cases}
0 & \text{if the }j^{\text{th}}\text{ rebound is on the left disk}, \\
1 & \text{if the }j^{\text{th}}\text{ rebound is on the right disk}.
\end{cases}
\end{align}
From $D$, he defined a number $A$ by its binary representation:
\begin{equation}
A = 0.d_1d_2d_3\dots = \sum_{j=1}^{\infty}2^{-j} d_j.
\end{equation}

\begin{figure}[h]
\includegraphics[width=15cm, angle=0]{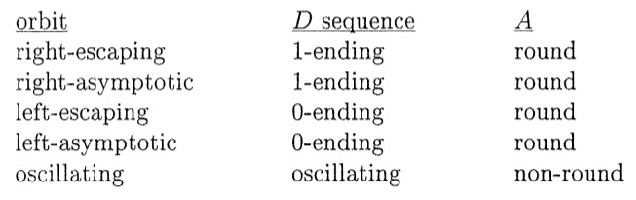}
\caption{Correspondance between the type of orbit, the $D$ sequence and the
  number $A$.}
\label{fig:d-a}
\end{figure}
He then proceeded to show that there is a close relation between the behaviour
of the $h$-orbits, $D$ and $A$, as given in Fig.~\ref{fig:d-a}.

Furthermore, for any given non-round $A$ there corresponds exactly one
$h$-orbit:
\begin{equation}
h = \left(\exp{\Phi} - 1\right) \sum_{j=1}^{\infty}2^{-j \Phi} S_j,
\end{equation}
where $s_j = -1$ if the $j^{\text{th}}$ rebound is on the left disk, and $s_j =
+1$ if the $j^{\text{th}}$ rebound is on the right disk.

In his usual way, {\MH} so an intriguing behaviour in a physical problem of
interest, found it amusing, and decided to study this behaviour for its own
right. He devised a model problem that allowed him to study the phenomenon in
its tiniest detail.

\section{Collisions and fragmentation}

The formation of asteroid families is the consequence of catastrophic impacts
on former parent bodies. Many workers have addressed the question of high
velocity collisions using numerical simulations taking into account both
experimental and theoretical results. But to reproduce the puzzling steep size
distributions of the asteroid families known at the time had been a task in
which these modelling techniques of fragmentation have typically failed.

In the late 1990's . Campo Bagatin and I addressed the problem from the point
of view of {\it geometrical constraints} (Campo Bagatin \& Petit, 2001). The idea that geometrical constraints
may play a role in the production of fragments was loosely found in the
literature since the 30's. We decided to address the problem in a detailed and
coherent way for the first time. Geometrical constraints stipulate that
fragments cannot overlap, and that putting together all the fragments must
reconstruct the parent body exactly.

Our main contribution to this study was to allow for arbitarilly given shape
for the largest fragment, with the other fragments being either triaxial
ellipsoids with apporximate axis ratios, or bodies of unconstrained shape. The
main goal here was to try and explain the observed steep slopes of the
cumulative size distributions of known asteroid families.  The geometric
effects were considered in realistic ways, and size distributions of the
produced debris were therefore obtained.

\begin{figure}[h]
\includegraphics[width=15cm, angle=0]{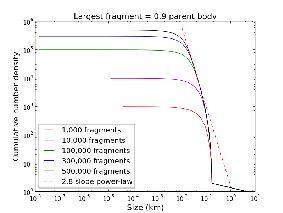}
\caption{Size distribution of spherical fragments with a largest remnant 0.9
  times the mass of the parent body. Different curves correspond to different
  total numbers of fragments simulated.}
\label{fig:frag}
\end{figure}

Running simulations with various numbers of fragments and sizes of the largest
remnant, we observed size distributions alike the ones observed for the
asteroid families. At the large-size end, the distributions exhibit a gap --
the bigger the larger the first remnant -- then a steep rise, followed by a
power-law regime and finally a plateau due to the finite number of fragments
(Fig.~\ref{fig:frag}). In the asteroid families, the (few) largest fragment(s)
seem to have a size and shape that is (are) stochastic, esentially related to
the past history of the parent body, and to the impact energy for the size of
the largest one. then we see a steep rise and a power-law like regime.
The power-law regime in our simulations was very satisfying, but it was unclear
were it came from, and even worst, we did not see how to determine the
exponent, except by measuring it on our results.

At the time, A. Campo Bagatin and I were working at Nice Observatory, in the
same corridor as {\MH}. Every day, we had lunch all together at the rather
famous Nice Observatory restaurant, enjoying its very fine cuisine while
overseeing Nice and the {\it Baie des Anges}. During the lunch, I frequently
had discussions with {\MH} about many things, including our own work. So one
day I mentioned to him our findings in the geometrically constrained
fragmentation simulations, in particular the appearance of a power-law regime.
He listened to me carefully, as he always did, and then we talked about other
things, and returned to our offices, each minding his own work, or so I
thought.

\begin{figure}[h]
\includegraphics[width=15cm, angle=0]{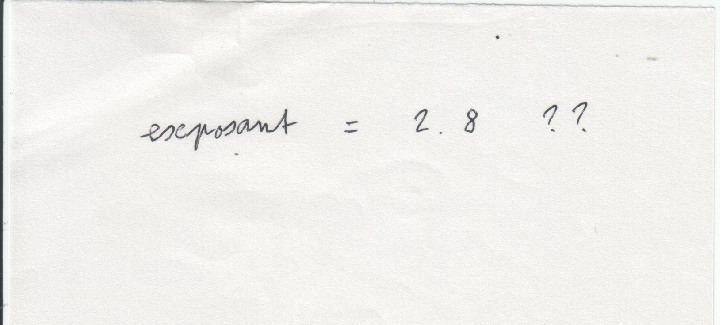}
\caption{Scanned version of the little piece of paper {\MH} brought to my
  office one day, giving the answer to the geometrically constrained
  fragmentation simulations.}
\label{fig:solution}
\end{figure}

A few days later, he showed up in my office, and gave me a little piece of
paper which I still keep dearly, and which is reproduced in
Fig.~\ref{fig:solution}. Since I only explained verbally the problem and did
not give him any of our results, his solution had to come from some anaytical
understanding of the process, not a measurement of the exponant from a
plot. And he could not resist the pleasure to exite my curiosity. Of course, he
was right, as can be seen from the dashed-line in Fig.~\ref{fig:frag}. When
pressed to explain himself, he gave me the outline of the reasoning, which not
only gave the answer we were looking for, but also gave a geenral formula
expressing the exponent $\alpha$ as a function of the dimensionality $a$ of the
problem:
\begin{equation}
\alpha = \frac{a^2 + 2 a -1}{a + 2}
\end{equation}
which yield $\alpha = 2.8$ for $a = 3$.

This result was very important for our work on geometrical constraints in
fragmentation simulations, as it gave us a full understanding of the
process. But unfortunately, it showed that the size distribution for the
smallest fragments (smaller than the thousand largest) is completely determined
by the algorithm, and hence bears no physical significance. We therefore had
restrict our study to the thousand largest fragments, and look at the effect of
the size and shape of the largest fragment: cratering case, spallation,
ellipsoidal core, conic antipodal fragment, ...

%

\section{\textbf{\textit{Le mot de la fin}}}

{\MH} was certainly the best possible advisor I could dream of. He gave me the
rigorous training I needed to embrass a research career. He reinforced in my
head the feeling that one must have a rigorous approach to research, as to any
other undertaking, trying to clearly define one's goal and context, and
consider all consequences of one's hypotheses, and also not be lured into
publishing before you have done a carefull and scientifically interesting
work. Even more important in some respects, he showed my that one must find the
fun in one's work, so as to keep interested and efficient at all times.
If one thing, this rigorous approach is what I try to apply to myself
every day in my work, and try to convey to my students.

\emph{Michel, You will stay in my mind and my heart for ever.}


\begin{thebibliography}{10}
\bibitem{cp01}Campo Bagatin, A. \& Petit, J-M., 2001, Effects of the Geometric Constraints on the Size Distributions of
  Debris in Asteroidal Fragmentation, Icarus, Vol. 149,p. 210-221.
\bibitem{h88a}H\'enon, M., 1988, Chaotic Scattering Modelled by an Inclined Billiard, Physica D, Vol.33, p. 132-156.
\bibitem{hp86a}H\'enon, M. \& Petit, J-M., 1986, Series expansions for encounter-type solutions of Hill's Problem, Celestial Mechanics, Vol.38, p. 67-100
\bibitem{ph86a} Petit, J-M. \& H\'enon, M., 1986, Satellite Encounters, Icarus, Vol. 66, p. 536-555.
\bibitem{ph87a} Petit, J-M. \& H\'enon, M., 1987, A numerical simulation of planetary rings. I. Binary Encounters., Astron. Astrophys., Vol. 173, p. 389-404.
\end{thebibliography}
\end{document}